# Crisis of Public Utility Deregulation
# And the Unrecognized Welfare State

Barbara A. Cherry

TPRC-2001-034

**Abstract.** Successful achievement of public policies requires satisfaction of conditions affecting political feasibility for policy adoption and maintenance as well as economic viability of the desired activity or enterprise. Fulfilling these joint requirements when pursuing deregulatory policies for public utility industries in the U.S. is particularly difficult given the legacy of the common law doctrines of "just price" and "businesses affected with a public interest." Prior research has discussed how deregulatory policies create economic pressures to increase reliance on rate rebalancing and explicit universal service funding mechanisms. Utilizing the framework developed in Cherry (2001), this paper examines the political feasibility problems encountered by attempts to address these regulatory changes. In this regard, it is helpful to view traditional public utility regulation as a form of welfare state regulation, where the obligation to provide essential services to all Americans has been delegated to private utilities. As with efforts to retrench from other forms of welfare state regulation, deregulatory policies create substantial political obstacles. The retrenchment problems are examined in the context of the electricity crisis in California as well as the passage and implementation of the Telecommunications Act of 1996. As expected, retrenchment from low residential retail rates – the most universalistic benefit for customers – faces the greatest political resistance. The political and economic circumstances of regulatory design are further complicated for the telecommunications industry given the pressures to expand or extend other dimensions of the welfare state, such as access to advanced services and the delivery of certain telecommunications services to educational institutions and health care providers. The societal trade-offs of monopoly as opposed to deregulatory policies must be reexamined in light of the instability created by the greater difficulties in satisfying both political feasibility and economic viability constraints under a deregulatory regime.

## 1. Introduction

Successful achievement of public policies requires satisfaction of conditions affecting political feasibility for policy adoption and maintenance as well as economic viability of the desired activity or enterprise. Fulfilling these joint requirements is becoming particularly difficult under deregulatory policies affecting public utility industries. The recent electricity crisis in California is a dramatic example of the adverse consequences – such as rolling blackouts, utility bankruptcies and government bailouts – that may arise when deregulatory policies conflict with conditions for economic viability. Fulfilling the joint requirements of political feasibility and economic viability in deregulatory policies for telecommunications is also becoming particularly difficult given the rapid rate of technological change, the growing complexities of communications technology, and the increasingly vital role of the information sector to global economies. In the U.S., the telecommunications landscape has been littered with competitive local exchange company bankruptcies, declining stock values and investments, and growing customer service problems. Throughout the European Union (EU), the telecommunications sector has shown similar distress, driven in large part by the excessive prices paid for third-generation mobile phone licenses.

To date, much of the academic research evaluating utility deregulatory policies has focused on designing regulatory incentives affecting behavior of private parties to better achieve desired policy goals. More recently, research has also emphasized the need to focus on the attributes of regulatory governance restraining the behavior of regulators in order to create a suitable environment for infrastructure investment (Levy & Spiller, 1996; Cherry & Wildman, 1999a).[1] Furthermore, Cherry and Wildman (1999a) have shown that the need to properly design both regulatory incentives and regulatory governance may require the sacrifice of some economic efficiency goals – in particular, short-term economic efficiency goals may need to yield to long-term ones – when transitioning from a monopoly to a competitive regulatory regime.[2]

The emphasis of prior research has been to encourage government officials – whether legislative, executive, administrative, or judicial – to better understand the constraints posed by the economic viability needs of firms and industries on public policy goals and associated regulatory designs. In so doing, many policy prescriptions have been made that appear, at least theoretically, to be quite straightforward. Yet many such policy prescriptions – for example, rebalancing retail rates and funding universal service through explicit charges on consumers' bills – tend to pose politically infeasible solutions (Cherry, 2000b; Cherry & Nystrom, 2000). *For this reason, it is important not only for policymakers to better understand the economic realities that limit achievability of policy goals, but also for all parties attempting to influence*

---

[1] Although Levy and Spiller (1996) and Cherry and Wildman (1999a) are written in the context of telecommunications deregulatory policies, many of the underlying economic and legal issues are similar and applicable to other public utilities.

[2] The author discusses the necessity of this trade-off in addressing stranded costs of incumbent local exchange carriers (ILECs) (Cherry & Wildman, 1999a) and pricing of unbundled elements by ILECs (Cherry, 2000a). More generally, the author also describes the economic constraints on regulatory interventions and the implications for applying constitutional principles to communications policies in the context of governance under the U.S. Constitution (Cherry & Wildman, 2000; Cherry & Nystrom, 2000).

*the policy process to be aware of the political constraints that limit policymakers' choices.* These political feasibility constraints may require modification or even abandonment of desired policy objectives or features of regulatory design to enable implementation of reasonably successful and sustainable telecommunications deregulatory policies.

Through incorporation of research from the political science literature, the author is attempting to fill this gap in the literature. Cherry (2001) provides the foundation for an analytical framework to design sustainable policies that meet both political feasibility and economic viability constraints.[3] This framework is based on recognition that political feasibility constraints on policy choices arise in three contexts: (1) to support the legitimacy of government itself; (2) to enable initial adoption of the policy; and (3) to enable sustainability of a policy over time. These contexts tend to emphasize differing types of political constraints as they address the policymaking process over differing time periods. In addition, policy choices must also simultaneously satisfy the economic viability constraints identified in Cherry and Wildman (1999a & 2000) that arise from the need: (1) to support private investment generally, and (2) to be compatible with the financial viability of specific firms or industries.[4]

Continuing along this line of research, this paper identifies specific political feasibility constraints in the U.S. that are impeding the adoption of sustainable policy objectives for the provision of public utility services under deregulatory regimes. Recent changes in technology and political philosophy have contributed to the pursuit of deregulatory policy objectives. Yet, deregulatory policies, necessarily requiring renegotiation of the terms under which public utility services are to be provided, are constrained by other longstanding policies arising from common law doctrines applied to public utilities in the U.S. More specifically, the prior policy choices embedded in the common law doctrines of "just price" and "businesses affected with a public interest" contribute to policy path dependence, creating inertia and daunting political obstacles to restructuring of retail prices in the electric and telecommunications industries.

Furthermore, this paper argues that the influence of these common law doctrines cannot be fully appreciated unless the policy choices embedded in public utility regulation are also understood in another political context.  In particular, public utility regulation in the U.S. can be viewed as a form of welfare state regulation, through which government seeks to provide access to certain essential services for the general population by delegating the task to private entities through a contractual (and later, a statutory) relationship.  As with other forms of welfare state regulation, retrenchment from traditional public utility regulation faces great political obstacles. This is particularly so for traditional policies of a universalistic nature, such as low retail rates for all residential customers.

---

[3] Cherry (2000b) identifies key economic and political factors underlying the divergence in rate rebalancing policies for telecommunications services adopted by the U.S. and EU.

[4] Although Cherry and Wildman (1999a & 2000) evaluate economic problems in the context of governance under the U.S. Constitution, the fundamental types of economic problems are similar across governance structures.

Recent changes in communications technology are also altering societal concepts of what constitutes essential telecommunications services – such as advanced telecommunications services – creating political pressures to expand what are considered to be "businesses affected with a public interest" to be provided at "just and reasonable rates."  Changes in communications technology also provide new means to deliver services traditionally recognized as part of the welfare state, such as education and health care.  These political pressures to extend social welfare are further complicating the ability to adopt and implement deregulatory policies.

This paper is organized as follows.  Section 2 describes the meaning and relevant history of the common law doctrines of "just price" and "businesses affected with a public interest" that are integral components of public utility regulation.  The legacy of these doctrines continues to constrain the policy options for revising public utility regulation.  Section 3 describes the concept of the welfare state and how public utility regulation in the U.S. can be viewed as a form of welfare state regulation.  In this regard, public utilities constitute privatized social welfare bureaucracies to ensure the provision of essential utility services to the general public.  Section 4 provides an overview of the framework in Cherry (2001) for designing sustainable policies that meet both political feasibility and economic viability constraints.  This section also discusses the unique political problems arising from attempts to retrench from preexisting welfare state policies through use of blame avoidance strategies.

Section 5, through integration of the discussion of the prior sections, analyzes the difficulties of retrenching from traditional public utility regulation by pursuing deregulatory policies.  In particular, deregulatory policies create economic pressures to increase reliance on rate rebalancing and explicit universal service funding mechanisms in a market open to competitive entry.  This section describes retrenchment problems associated with rate rebalancing – particularly from the traditional policy of low retail rates for residential customers

– that have contributed to the development of the electricity crisis in California and to the codification of restrictions on rate restructuring in TA96. This section also describes the retrenchment problems of rate rebalancing and explicit universal service funding mechanisms, as well as the political pressures to expand aspects of the welfare state, that are creating obstacles for implementing universal service policy in telecommunications. Section 6 provides a summary and conclusions.

## 2. Common Law Doctrines Underlying Public Utility Regulation

In the U.S., public utilities constitute a specific, legal category of entities that evolved under the common law. In essence, public utility law developed from the confluence of other developments in the law related to governments' power to regulate commercial activities, governments' authority to delegate its powers to private parties through franchises, and businesses' ability to organize their activities into a corporate form (Cherry, 1999, pp. 50-57).

As discussed more fully in this section, some of the legal obligations imposed on public utilities derive from common law doctrines that had originally been applied to a wider scope of commercial activities. Two of these doctrines – a deeper understanding of which is essential in order to appreciate the political difficulties in altering their application under deregulatory policies – are "just price" and "businesses affected with a public interest". As a result, under the common law (and subsequently reflected in many states' statutes) public utilities have the obligation to provide adequate service and facilities at just and reasonable rates and to render service to all applicants without discrimination.

However, because privately-owned public utilities have also been granted some form of privilege from government to be exercised for the benefit of the public – such as eminent domain powers – historically the government has imposed greater obligations on the economic activities of public utilities as compared to general businesses. These include an affirmative obligation to serve a specified territory that requires expansion of facilities and capacity, even at non-compensatory rates, and a prohibition from withdrawing service without prior government approval. This latter set of obligations arose from what was originally a contractual relationship with government, usually based on monopoly franchise agreements with local units of government. Eventually this relationship became codified through federal and/or state legislation. The importance of this "contractual" relationship[5] between public utilities and government will be discussed in Section 3.

### 2.1 Legacy of the Just Price Doctrine

The origins of the American concepts of "just and reasonable rates" and "business affected with a public interest" can be traced to the medieval doctrine of the "just price". The just price doctrine was a legal device to enforce a specific moral imperative of justice on economic transactions. It evolved in Europe during the period of Scholasticism in economic thought through the integration of Aristotelian philosophy, Roman law and canon law.[6]

The just price doctrine required the equivalence of value in exchange so that the price reflects the value of the good or service for the community in general, and is not excessively high or low due to unique circumstances of specific sellers or buyers (Langholm, 1992; pp. 181-189; Baldwin, 1959, pp. 71-80). This requirement was based on the concept of commutative justice – whereby each party is given his due – to address situations of economic coercion, exploitation, and the illegal wielding of bargaining power. As a consequence, either buyers or sellers could seek restitution with regard to an unjust price. It bears stressing that the just price doctrine was intended to address ethical, moral problems regarding the use and misuse of bargaining power in order to protect persons from exploitation. Persons deemed in need of protection included not only the poor with respect to essential commodities but also any person under economic compulsion – that is, any person isolated from normal economic relations that deprive them of the bargaining power necessary to withstand exploitation (Langholm, 1992, pp. 578-579).

The just price doctrine was a significant component contributing to the development of the concept of "public callings" under the English common law during approximately the fourteenth century.[7] At that time the principle social and legal institution was the feudal relation of lord to man, with rights and duties based on that relation. Performing services in the context of this relation was considered private employment; however, rendering services outside of this relation to the public was considered public employment. Public employments included not only common carriers or innkeepers, but also other occupations so long as they were provided to the public, such as blacksmiths and surgeons. Public callings were simply undertakings to serve the public, and by virtue of their *status as public employments* various legal duties or obligations were imposed under the common law that did not apply to private employments. In particular, those engaged in public callings were required to charge reasonable prices, to serve without discrimination, and to exercise their calling with adequate care, skill and honesty.

---

[5] A franchise agreement between the public utility and a unit of government is a legally enforceable contract. When the franchise agreements are superseded by statute, the legal relationship between the public utility and government is usually no longer considered one based on contract (Cherry, 1999, pp. 30-32, 50-57; Cherry & Wildman, 2000, pp. 81-85). This is sometimes the source of confusion, as economists continue to refer to the public utility relationship with government as a contractual one (Goldberg, 1980).

[6] For a detailed discussion of the origins of the just price doctrine, *see* Baldwin (1959), Langholm (1992 & 1998) and Schumpeter (1994).

[7] For discussion of the development of public callings, *see* Glaeser (1957) and Adler (1914).

Importantly, the obligations of public callings were based on tort, not contract, law. In other words, public callings bore these obligations merely by engaging in the activity not because the parties to the transaction agreed to such obligations in the context of negotiating some form of agreement.

### 2.2 And the Development of "Business Affected With a Public Interest"

With the rise of mercantilism, feudalism as a social and legal institution declined and, correspondingly, public employments came to dominate private employments for conducting economic transactions. Analogously, claims for damages arising from economic transactions gradually shifted from a basis in tort to one in contract with the evolution of a more sophisticated common law of contracts. (Glaeser, 1957, pp. 200-202; Horwitz, 1977; Cherry, 1999, pp. 47-51) With the rise of laissez faire ideals, the meaning of public callings continued to wane during the seventeenth and eighteenth centuries so that, by the early nineteenth century in the U.S., most general businesses came to be regulated under the obligations of contract law. In other words, for general businesses the courts' role of providing protection from unequal bargaining power eroded and increased reliance was placed on the freedom to contract. As a result, the medieval obligations of public callings no longer applied to general businesses.

*However, the tort obligations of public callings were retained for "business affected with a public interest."* Businesses affected with a public interest were those for which it was deemed that the dependence of the customer required protection. Such businesses fell into three categories:

1. Those which are carried on under the authority of a public grant of privileges which either expressly or impliedly imposes the affirmative duty of rendering a public service demanded by any member of the public. Such are the railroads, other common carriers and public utilities.
2. Certain occupations, regarded as exceptional, the public interest attaching to which, recognized from earliest times, has survived the period of arbitrary laws by Parliament or colonial legislatures for regulating all trades and callings. Such are those of the keepers of inns, cabs, and gristmills.

3. Businesses which, though not public at their inception, may be fairly said to have risen to be such, and have become subject in consequence to some government regulation. They have come to hold such a peculiar relation to the public that this is superimposed upon them.

*(Wolff Packing Co. v. Court of Industrial Relations of Kansas* (1923), p. 535, citations omitted).

From examination of the case law, attributes in common among these three categories are: (1) that the service is of special public importance or necessity; (1) that circumstances or characteristics of supply are such that the service is not available in a competitive market; and (3) that the activity has current and/or future widespread effects on the community at large. Significantly, the second attribute includes numerous situations in which competition is considered impracticable. One might be the grant of some special governmental privilege, perhaps even a legal monopoly, as with common carriers and public utilities. However, it also included situations of "virtual monopoly" that arise without government involvement, such as firms being strategically situated in terms of location (grain elevators) or time (innkeepers with respect to travelers).[8]

The inherent power of government to regulate is referred to as the police power. An important function of the courts is to determine when government regulation falls within the scope of the police power without violating provisions of the U.S. Constitution. The significance of identifying certain businesses as being "affected with a public interest" is that the scope of permissible regulation under the police power was deemed to be greater than that for other general businesses. During the nineteenth century, the permissible scope of regulation of businesses *not* affected with a public interest was quite narrow, so that the validity of government regulation often hinged on governments' ability to convince the courts that the business in question did fall into one of the classes of business affected with a public interest.

In the famous case *Nebbia v. New York* (1934), the U.S. Supreme Court effectively broadened the scope of permissible regulation under the policy power *for any business* by declaring that "it is clear that there is no closed class or category of businesses affected with a public interest" and that " 'affected with a public interest' is the equivalent of 'subject to the exercise of the policy power' "(p. 535).  Given that the scope of the policy power was deemed synonymous with regulation in the public interest, the need to prove that a business did or did not fall into the historical classes of business affected with a public interest fell into disuse.

---

[8] In fact, in the landmark case of *Munn v. Illinois* (1934), the grain elevator at issue satisfied this criterion because the "elevator was strategically situated and … a large portion of the public found it highly inconvenient to deal with others" (*Nebbia v. New York*, 1934, p. 532). It should also be noted that the grain elevator fell into the third category of businesses affected with a public interest described above.

As a result, most commentators claim that the traditional definition of "business affected with a public interest" is now totally irrelevant.[9]  However, this is a misreading of *Nebbia v. New York*.  Although the Court found that the police power was coextensive with regulation in the public interest, it still maintained that permissible regulation as to a *given business* depends on the specific circumstances in each case.

> It is clear that there is no closed class or category of businesses affected with a public interest, and the function of courts in the application of the Fifth and Fourteenth Amendments is to determine in each case whether circumstances vindicate the challenged regulation as a reasonable exertion of governmental authority or condemn it as arbitrary or discriminatory. (p. 536)

In fact, the Court proceeded to examine the specific circumstances of the case – concerning the constitutionality of price regulation of retail milk sales in New York – in a manner reflective of the traditional attributes of businesses affected with a public interest.  The Court found that the dairy industry, having not received any public grant or franchise, was clearly not a public utility, nor was it a monopoly.  However, state legislative investigation in 1932 "was persuasive of the fact that … unrestricted competition aggravated existing evils, and the normal law of supply and demand was insufficient to correct maladjustments detrimental to the community" (p. 530), and "in these circumstances the legislature might reasonably consider further regulation and control desirable for protection of the industry and the consuming public" (p. 530).  In essence, the circumstances that the Court found to be compelling were the importance of the product, maladjustments of the market, and widespread impact on the community – reminiscent of the traditional attributes one through three, respectively.

Given the holding and analysis in *Nebbia v. New York*, a wider range of businesses can now be subject to some government regulation.  However, what would be deemed a reasonable exertion of that governmental authority is still likely to be greater for a business in which the circumstances are similar to those of the traditional justifications for regulating "businesses with a public interest."

---

[9] *See, e.g.* Barnes (1942, pp. 12-13) and Bonbright (1961, pp. 5-7).  However, Grieve & Levin (1996) also stresses the continuing importance of the legal public utility concept in a competitive public utility market structure.

### 2.3 Codification of Public Utility Regulation

Yet, the continuing importance of the traditional definition of "business affected with a public interest" is particularly acute for public utilities and some common carriers. This is because for most of them the common law obligations to charge reasonable prices, to serve without discrimination, and to exercise their business with adequate care and skill have been codified in legislation.

As public utilities, the electric and telecommunications industries are regulated by state laws that are enforced by state commissions. With regard to interstate commerce, the electric and telecommunications industries are regulated by federal statutes and enforced by the Federal Energy Regulatory Commission (FERC) and the Federal Communications Commission (FCC), respectively. In both industries the medieval obligations of public callings (and subsequently, of traditional "businesses affected with a public interest") described earlier have been expressly codified in statutory language. In particular, the just price obligation has been consistently codified through use of the phrase "just and reasonable rates".

Once enacted, these laws can be changed only by subsequent legislation or through invalidation by judicial litigation. As just previously discussed, invalidation by the courts is unlikely given the legacy of common law and constitutional jurisprudence. In fact, if the laws were simply repealed (without enactment of other substantive statutory provisions) the same duties would likely be reimposed by virtue of the common law.

Furthermore, adoption of legislation to repeal or modify common carrier or public utility law is itself a difficult process. This is due to the general tendency for inertia and policy path dependence associated with the policymaking process, especially for decentralized and fragmented governance structures such as those in the U.S. (Wilsford, 1994). It is noteworthy in this regard that the Communications Act of 1934 remained unaltered until the passage Telecommunications Act of 1996 – and even that occurred only after seven years of intense lobbying and a shift in party power in Congress after the 1994 elections.

More critically, as will be discussed in the following sections, attempts to modify existing law through pursuit of retrenchment policies – which include attempts to retrench from the policy of "just and reasonable rates" for retail utility customers – are particularly hazardous for politicians.  It is in this political context that the legacy of public utility regulation, and notably the just price doctrine, has its greatest influence.

## 3. Public Utility Law as Welfare State Regulation

Political difficulties in adopting deregulation policies for public utilities can be better appreciated by also understanding the development of another important social and legal institution, namely, the welfare state.  This section will describe the concept of the welfare state and how public utility regulation in the U.S. can be viewed as a form of welfare state regulation.  The political ramifications of recognizing public utility regulation as a form of welfare state regulation will be discussed in section 4.

### 3.1 Defining the Welfare State

There does not appear to be one precise, generally agreed upon definition of the welfare state.  Yet various authors have offered their articulations of what constitutes the welfare state.  Wolfe (1989, pp. 107-09) asserts that the welfare state exists when government assumes the task of protecting the moral order that makes society possible.  Pinch (1997, p. 150) defines the welfare state as a "set of institutions and social arrangements designed to assist people when they are in need because of factors such as illness, unemployment and dependency, through youth and old age."  Mishra (1990, p. 34) defines the welfare state as "the institutionalization of government responsibility for maintaining national minimum standards."  Cranston (1985, p. 4) acknowledges that for some interpretations the essence of the welfare state is "government-protected minimum standards of income, nutrition, health, housing and education, assured to every citizen as a political right, not as charity", but other interpretations have identified additional aspects, such as social welfare, fiscal welfare and occupational welfare.[10]

Common elements of these varying definitions and interpretations are the institutionalization of some minimum level of rights or access to essential goods and services through government intervention.  Importantly, these elements represent a particular concern with distributive justice – distributing societal rights and resources in light of personal worth and need.  This is in contrast to the concept of commutative justice underlying the just price doctrine (Baldwin, 1959).

---

[10] Fiscal welfare refers to the benefits available through the tax system such as allowances for expenditure.  Occupational welfare consists of work-related benefits such as sick pay, paid holidays, pension schemes, health and safety conditions, and subsidies for housing, recreation, meals and education. (Cranston, 1985, p. 4)  The elements of social welfare are described later in this section.

The present discussion focuses on social welfare, which is the most characteristic part of the welfare state (Cranston, 1985). Cranston provides a useful typology of social welfare, characterizing techniques of social welfare into two categories. The first is *social welfare regulation,* which consists of "the legal regulation of private institutions such as property owners and employers in the interests, at least ostensibly, of those such as tenants and employers" (1985, p. 5). Both beneficiaries and the private parties on whom mandates of behavior are placed are identified by their status, such as tenants and employees or property owners and employers, respectively.[11] Examples of social welfare regulation include rent control and housing standards imposed on property owners, and minimum wage and OSHA laws imposed on employers.

The second category of social welfare is *social welfare benefits and services* (Cranston, 1985, pp. 4-5). This category consists of the imposition of duties placed on social welfare bureaucracies to provide benefits and services to individuals in specific circumstances. This category is then often subdivided into two groups. One consists of benefits and services providing a minimum standard, such as food stamps and public housing. Because eligibility is unearned and means-tested, these benefits and services are considered *residualistic* in nature. The other group consists of benefits and services provided to a much broader set of beneficiaries, such as public education and social security in the U.S. and health care in the UK. Given that eligibility is considered earned, or at least not means-tested, these benefits and services are considered *universalistic.* Although it is not always clear as to which group a given benefit or service should be assigned, the distinction is often very useful. As discussed in Section 4, this is particularly true for understanding the differing political realities facing adoption and maintenance of residualistic as compared to universalistic benefits and services.

### 3.2 Public Utilities as Social Welfare

Although some interpretations of social welfare have included a broader set of legal interventions, such as fiscal or occupational welfare,[12] none have included public utility regulation. Such, probably unwitting, exclusion has overlooked the existence of an important form of regulation imposed under the American common law that bears the same characteristics of other previously recognized forms of social welfare. As previously discussed, the common law doctrine of "business affected with a public interest" imposed specific behavioral requirements by virtue of a business' status. Such businesses bore special obligations because of the special public importance and necessity of the service they provided, circumstances or characteristics of supply that prevented availability of the service in a competitive market, and the widespread effects of their activities on the community at large. These characteristics are similar to other forms status-based regulation, such as behavioral mandates on property owners and employers, which have already been recognized as social welfare regulation.

---

[11] It is worth noting that assignment of duties or obligations based on status is a characteristic of tort law, from which the just price doctrine was derived.

[12] See note 10, *supra.*

Furthermore, as the common law evolved, the duties of public utilities arose from a combination of their status as businesses affected with a public interest and their contractual relationships with government under which certain affirmative obligations to serve were imposed. Belonging to one of the traditional classes of businesses affected with a public interest, public utility services were deemed of special importance and necessity to the community. But, unlike other classes of businesses affected with a public interest, public utilities were granted special privileges from government to conduct their business and in return bore additional special obligations to the communities they served. It is this dual, contractual and status-based nature of public utility regulation that provides characteristics similar to those services previously recognized as belonging to the second category of social welfare benefits and services. However, there is one interesting difference. Social welfare bureaucracies have implicitly been assumed to be government entities. In the case of public utilities, the duties were delegated to private bureaucracies – public utilities – pursuant to a contractual relationship with government. Over time this contractual relationship was converted to a statutory one through legislation, under which public utility services continue to be administered through private firms but with oversight by federal and/or state commissions.

Viewed as a form of social welfare service, traditional public utility regulation is primarily of the universalistic form. This is because the public utility duties – embodying the legacy of the just price doctrine – are imposed for the benefit of all customers in its serving area. All customers are to receive service at just and reasonable rates, without discrimination, and under an adequate standard of care. Furthermore, the public utility must stand ready to extend facilities to any customer in the relevant serving area, even under non-compensatory rates. Under historical price structures, residential customers have usually benefited from a lower level of "just and reasonable" rates relative to business customers for the same services, often requiring some form of cross subsidy to sustain the lower level of residential rates. It is in this respect that public utility services are universalistic with regard to residential customers. However, some classes of residential customers or services have also been subsidized by other classes of residential customers or services, such as rural by urban or local by toll with regard to telecommunications services. In this respect, some aspects of public utility regulation take on more residualistic forms. As will be discussed in section 5, deregulatory policies give rise to economic viability problems that create pressure to convert even more aspects of public utility regulation to a residualistic form.

For the preceding reasons, traditional businesses affected with a public interest and public utility regulation can be viewed as forms of social welfare regulation.[13] However, whether or not public utility regulation is formally recognized as an element of the welfare state is not what is critical. What is important is that public utility regulation shares the same characteristics as previously recognized forms of welfare state regulation from which retrenchment poses similar political problems.

## 4. The Political-Economic Dynamics of Adopting and Maintaining Sustainable Policies

Policy choices that are likely to lead to fulfillment of the underlying policy objectives are constrained by political feasibility, as well as economic viability, problems. It is beyond the scope (and page limit) of this paper to fully reiterate the analysis provided in Cherry and Wildman (2000) with regard to economic viability constraints. Nor is it possible to recount in full detail the analysis in Cherry (2001), which provides an analytical framework for understanding general political feasibility problems and their interrelationship with economic viability constraints. However, an overview of the critical elements and conclusions of these analyses are identified and briefly discussed in this section.

The following subsections expand upon these basic points. In order to be economically sustainable, policy choices affecting activities of private firms face two basic types of economic viability constraints. One is the need for the policy to support private investment as a general matter. The second is the need for the policy to be compatible with the financial viability of the specific firms or industries engaged in the activity affected by government regulation. For political sustainability, policy choices face political feasibility constraints in three different, albeit interrelated, contexts: (1) to support the legitimacy of government itself, (2) to enable initial adoption of the policy through the policymaking process, and (3) to enable the policy to remain in force over time. Simultaneously satisfying all of these economic and political constraints is often a difficult endeavor. This is because the constraints are not merely additive, but interrelated and often incompatible. For example, as discussed more fully in Section 5, some features of deregulatory policies – such as freezing of retail but not wholesale electricity rates in California – may enhance political feasibility but undermine economic viability.

---

[13] Interestingly, if these common law obligations are expressly recognized as creatures of the welfare state, then it would provide a counterargument to the common assertions that the U.S. has been a laggard in the development of a welfare state.

### 4.1 Economic Viability Constraints on Policy Choices

To achieve desirable policy objectives, the economic realities of providing goods and services through private entities places constraints on the design of governmentally imposed regulatory rules, both on private parties through regulatory incentives and on government entities through regulatory governance. If these economic realities are not addressed, the desired economic performance and social consequences underlying policymakers' objectives may not be forthcoming. The basic types of economic constraints are briefly discussed here.

First, government's own performance influences what can be achieved by private entities in a system of voluntary exchange, because it affects the long-term certainty and risk that parties face. Government contributes to the viability of the market itself through definition and enforcement of private property rights and rules of contract. However, these rules must constrain government as well as private party behavior. Such enforceability is necessary in order to ensure that government can, in fact, make credible commitments and thereby preserve its capacity to make contracts in the future. It is also necessary to support the general system of private property rights. Both of these attributes are critical to support private investment in utility infrastructures (Levy & Spiller, 1996; Cherry & Wildman, 1999a).

Second, even if a system of regulatory rules generally supports private investment in the economy, rules applied to a specific sector or industry may not be compatible with the economic viability of the affected firms or industries. Regulatory rules may pose economic viability problems for a given firm or industry, among firms within a given industry, or among industries. These problems can be designated as prospective or transition problems. *Prospective problems* arise from the prospective effects of government rules that (1) treat some firms of industries differently than others, whether on a per se or de facto basis; (2) impose unreasonable and fundamentally unremunerative financial obligations on firms or industries; or (3) require compliance with coexisting yet conflicting or incompatible rules. *Transition problems* arise from changes in governmental rules that affect the earnings on preexisting investments, contracts, or conduct, and thereby the willingness of private actors to rely on government commitments in planning future economic endeavors.

Failure to address the relevant types of economic viability constraints can lead to the failure of achieving desired policy objectives. In some circumstances, pursuit of policies without adequately addressing these constraints can create significant adverse consequences, even yielding net societal losses. This latter point is discussed further in Section 5.

### *4.2 Political Viability Constraints on Policy Choices*

#### *4.2.1 To Support Legitimacy of Government Itself*

Successful pursuit of policy objectives requires, perhaps most fundamentally, that regulatory intervention be constrained by those limitations on government action that support the legitimacy of government itself. This is particularly true of democracies (Habermas, 1999, p. xxv (translator's introduction)). Analogous to the need for government to constrain its own behavior in order to generally support private investment in a market economy, government must also adhere to the terms of its "social contract" reflected in its governance structure in order to maintain credible commitments to rule (Black, 1993; Hoepfl & Thompson, 1979).

#### *4.2.2 To Enable Adoption of a Policy*

A policy choice is also constrained by the circumstances prevailing at the time of its adoption. These constraints are endemic to the policy decision-making process itself. Most public policy change is incremental and major policy change requires the intervention of strong conjunctural forces (Hall, 1986; Wilsford, 1994). In this way, the policy process is characterized by a high degree of path dependence and inertia. The strength of the conjunctural force necessary for a major policy change varies, of course, among governance structures, and thus so does the degree of path dependence and inertia. An important observation for considering deregulatory policies is that greater inertia is associated with highly fragmented political institutions, such as the U.S., than with more strongly centralized state structures, such as Britain and Germany (Wilsford, 1994).

Cherry (2001) utilizes Kingdon's model (1995) of the policymaking process to more fully describe the political feasibility problems associated with seeking adoption of a policy choice at a given point in time. Page limitations preclude explication of this model here for purposes of providing an admittedly more analytically satisfactory accounting of the political feasibility problems. Therefore, the reader is necessarily referred to Kingdom (1995) and Cherry (2000b; 2001) for a description and application of this model. But, for purposes of the discussion in Section 5, the salient points are that: (1) the policymakers' views of economic viability problems control the policy agenda; and (2) the policymakers' views of political feasibility ultimately determine both the attributes of the policy choice to be pursued and the political strategy deemed necessary for its adoption.

As to the first point, policymakers' views of economic viability problems are affected by their perceptions of prior policy choices. This is just one of the elements of path dependence in the policy decision-making process. Their perceptions of policy problems are influenced by various information sources – including external ones such as industry members, lobbyists, mass media and policy experts – providing a wide range of often-conflicting perspectives. Policymakers also evaluate the relative importance of perceived economic viability problems in light of other policy problems requiring their attention. The realities of limited time and resources compels policymakers to compare and rank the importance among numerous, often unrelated policy problems – a task often overlooked, under appreciated or further confused by those sources of information upon which policymakers rely.

These factors contribute to the likelihood that there will be gaps among policymakers' views of economic viability problems and the views of those attempting to influence them. Importantly, one type of gap may likely be between policymakers and policy experts, where the theoretical assumptions underlying expert opinions may fail to adequately incorporate political realities that only serve to widen the gap.

As to the second point, policymakers' views of the appropriate policy option and the political strategy to pursue it is driven by what is perceived to be politically possible under existing circumstances. In this regard, the process of successful coalition building is essential. Policymakers must evaluate the organization of political forces in support or opposition to the policy option, perceived public opinion, and the existence of other policymaker approval. In turn, these evaluations will be affected by prior experience under similar circumstances.

Furthermore, and perhaps more importantly, these evaluations will be made in the context of policymakers' own political objectives. For policymakers driven by electoral motivations, the choice of political strategy is determined by several factors: constituent gains and losses; the concentration versus diffusion of costs and benefits among constituents; and the negativity bias of voters (i.e. constituents respond more to losses than gains), causing politicians to discount gains relative to losses (Weaver, 1986, pp. 373-374).

Upon evaluation of these factors, policymakers choose among credit claiming strategies or blame avoidance strategies. With *credit claiming strategies*, policymakers are willing to communicate to constituents and the media the attributes of the proposed policy, the intended beneficiaries, and the likely consequences if the policy is adopted. In fact, policymakers rely on such communication to reap the desired electoral rewards. For this reason, credit claiming strategies are favored when a policy option produces benefits to a broadly defined set of constituents, or to a group of concentrated constituents but with diffusely distributed constituent losses to inhibit the development of organized forces of political opposition. (Weaver, 1986; Twight, 1991).

In contrast to credit claiming strategies, policymakers favor *blame avoidance strategies* when a policy option requires retrenchment of substantial benefits from a concentrated group of constituents but confers relatively small benefits to a diffuse group of constituents. This is because such a distribution of losses and benefits imposes significantly lower transaction costs to organize political opponents rather than supporters. In addition, the electoral hazards are intensified by the negativity bias of voters (Pierson, 1994; Weaver, 1986). Blame avoidance strategies consist of distinctive tactics to diffuse political opposition (Pierson, 1994, p. 8). They include obfuscatory tactics to decouple the relationship between the desired policy and its negative consequences through manipulation of information available to constituents; avoidance of deciding critical policy elements through delegation to other governmental entities; and compensation to victims of retrenchment, such as grandfather clauses (Pierson, 1994; pp. 19-26; Weaver, 1986, pp. 384-390). Reliance on blame avoidance strategies – reflecting policymakers' fears to dismantle existing policy and thereby limiting the perceived feasible set of policy alternatives – contributes to the path dependency of preexisting, even failing, policies (Weaver, 1986, pp. 393-395).

As Pierson (1994) explains in great detail, policy retrenchment is a distinctive and difficult political enterprise. Describing welfare state retrenchment efforts in Great Britain and the U.S. during the 1980's, he states:

> …[Retrenchment] is in no sense a simple mirror image of welfare state expansion, in which actors translate … a favorable balance of class "power resources" or institutional advantages into political success. Retrenchment advocates must operate on a terrain that the welfare state itself has fundamentally transformed.

Welfare states have created their own constituencies. If citizens dislike paying taxes, they nonetheless remain fiercely attached to public social provisions. That social programs provide concentrated and direct benefits while imposing diffuse and often indirect costs is an important source of their continuing political viability. Voters' tendency to react more strongly to losses than to equivalent gains also gives these programs strength.

Retrenchment advocates thus have their work cut out for them. Almost always, retrenchment is an exercise in blame avoidance rather than credit claiming. Even a government like Margaret Thatcher's, possessing centralized political authority and confronting a weak and divided opposition, had to acknowledge the potential for widespread popular disapproval of significant reforms. The Reagan administration, operating from what was in most respects a weaker institutional position, faced even greater difficulties. (1994, pp. 1-2; footnote omitted)

As will be discussed in Section 5, deregulatory policies for public utilities are retrenchment policies. As such, they required the use of blame avoidance strategies and are particularly difficult to adopt. Furthermore, as evidenced by the experience in California, even if adopted - however flawed – they may also be difficult to change.

*4.2.3 To Enable Sustainability of a Policy Over Time*

Even if a policy choice is initially adopted, fulfillment of the underlying policy objectives may require the sustainability of that policy over time. Of course, the relevant time frame that a given policy needs to remain in effect depends upon the specific policy objectives and economic conditions required to achieve them. Section 4.2.2 described political feasibility problems associated with adoption of a policy at a given point in time. However, the ability to retain a policy over time requires analysis of the political problems associated with surviving subsequent efforts of retrenchment. This requires a dynamic, not static, assessment of the policy decision-making process over time.

Political scientists have examined the characteristics of policies that tend to better withstand attacks of retrenchment. Perhaps most relevant to the consideration of adopting sustainable public utility deregulatory policies are the conclusions of research in the context of social welfare programs. In democracies, universalistic programs are more politically sustainable than residualistic ones (Mishra, 1990; Skocpol, 1995; Wilson, 1987). The underlying reason is that the more broadly defined the group of beneficiaries, the broader the support from constituencies for maintaining the existing policy notwithstanding changes in circumstances since their adoption.  For this reason, universalistic programs are more politically sustainable even if they are more expensive than policies targeted solely on the poor or marginal groups (Skocpol, 1995, pp. 250-253). [14]  Consequently, some political scientists advocate "targeting within universalism", that is, addressing the needs of the less privileged through programs that include more advantaged groups (Skocpol, 1995, pp. 267-272; Wilson, 1987, p. 1188-124).  This recommendation is in stark contrast to those of many economists who advocate, for example, narrowly targeted universal service programs as a component of telecommunications deregulatory policy in order to minimize the funding burden.

The importance of the discussion here is that factors affecting the political sustainability of a given policy option *over time* also need to be contemplated when designing and selecting a policy option for adoption *at a given point in time*.  In other words, incorporation of the factors of Section 4.2.3 into the analysis of Section 4.2.2 raises the likelihood of adopting a policy option that actually fulfills the desired policy objectives.  Of course, the available options remain constrained by the overall set of options supportive of the legitimacy of government itself, as described in Section 4.2.1.  In this way, compliance with all the political feasibility constraints described in Section 4.2 must be achieved simultaneously.

**5. Crisis of Public Utility Deregulatory Policies**

An essential element of deregulatory policies is the elimination of legal monopolies for public utilities and the opening of market entry.  This change in market structure poses challenges for sustaining the traditional policy goals of "just and reasonable prices" and the widespread availability of public utility services.  To meet economic viability constraints, pursuit of these goals requires increased reliance on rate rebalancing and explicit funding mechanisms for universal service objectives.  Recognizing how traditional public utility regulation shares distinguishing characteristics of other recognized forms of social welfare benefits and services enables one to appreciate how pursuit of rate rebalancing and explicit funding mechanisms pose similar political problems of policy retrenchment.

---

[14] Skocpol (2000) also identifies other characteristics associated with successful social policy programs in the U.S.  These are: (1) benefits provided in exchange for service rather than as entitlements; (2) policies nurtured by partnerships between government and popularly-rooted voluntary associations; and (3) programs backed by reliable public revenues.  The validity of these characteristics may vary among nations, depending upon, among other things, differences among their institutional endowments (Levy & Spiller, 1996).

This section describes retrenchment problems associated with the electricity crisis in California as well as passage and implementation of TA96. As expected, attempts to raise residential retail rates – threatening the universalistic benefit – face perhaps the greatest resistance. Problems also arise from the rural telecommunications carriers' efforts to protect vested interests in preexisting high cost funding, and the blame avoidance strategies employed by politicians to avoid accountability for the "tax" imposed on carriers to fund explicit funding mechanisms. The economic pressures from the funding mechanisms are also exacerbated by other policy objectives to expand the public utility welfare state to advanced telecommunications services and to extend the availability of more traditional forms of welfare state services – that is, education and health care services – through delivery by more advanced telecommunications technology.

The political problems facing deregulatory policies for public utilities are consistent with those experienced with efforts to retrench from other forms of welfare state regulation. As a general matter, the more universalistic benefits will continue to pose the greatest resistance to retrenchment. However, the economic viability problems are more complex, for continued provision of public utility service relies on the continued financial viability of privatized firms in the industry rather than obtaining more funding for a public bureaucracy from some government budget. Simultaneously satisfying the economic viability and political feasibility constraints of deregulatory policies over time will be particularly difficult as circumstances change. The recent electricity crisis in California is but one of what will likely be a long list of examples. By contrast, being monopoly-based, traditional public utility regulation has been both politically and economically more stable.

## 5.1 Retrenchment Problems of the Public Utility Welfare State

It is widely recognized that, in forcing a transition from regulated public utility monopolies to a more competitive market structure, deregulatory policies inevitably create economic pressures to alter pricing structures of public utility services. In order to maintain universal service-related policy objectives under a competitive market structure, policymakers must shift reliance from subsidies embedded in the traditional price structure to some combination of rate rebalancing and explicit funding mechanisms (Cherry & Wildman, 1999b). This policy shift toward rate rebalancing and explicit funding mechanisms poses great political hazards for deregulatory policy advocates.

*5.1.1 Retrenchment Problems of Rate Rebalancing*

First, rate rebalancing creates pressure to move from universalistic to residualistic policies, or to make existing residualistic programs even more narrowly targeted. The economic realities of open entry and the development of niche market players are that price levels will be driven closer to costs and will become less uniform in order to target groups of customers to reflect varying usage needs. These economic realities render subsidies of the traditional public utility rate structures unsustainable in the long run.[15] Business rates can no longer contribute the same levels of revenues (relative to costs) toward the maintenance of lower residential rates. Targeting of customers according to usage patterns and geographic locations undermines efforts to maintain geographical subsidies, such as urban to rural residential telecommunications customers, or rate averaging.

Those in the varying groups benefiting from low rates under pressure to rise will suffer losses, and would be expected to oppose such restructuring. This is the source of retrenchment problems for public utility customers. For example, given the experience from other retrenchment efforts discussed in Section 4, the universalistic nature of traditionally low rates for residential customers should pose great political difficulty for the adoption of policies that would require significant increases in the overall rate levels of residential customers. For the more residualistic aspects of public utility pricing, those users facing potential losses would also be expected to resist retrenchment efforts. For example, among telecommunications residential customers one would expect opposition from rural (relative to urban) and local (especially those with low relative to high toll usage) customers.[16]

---

[15] There has been considerable debate regarding what can be properly described as providing a "cross subsidy," which usually focuses on disagreements regarding the allocation of fixed costs of the network among services. In the present context, the term subsidy does not necessarily imply that some classes of services or customers are subsidizing others in the strict economic sense, but that government intervention requires some services to recover costs that would otherwise not be sustainable in a competitive, unregulated environment (Panzar & Wildman, 1995).

[16] There may also be retrenchment problems for public utility service providers, depending upon whether the deregulatory policies are perceived to be imposing net losses relative to the status quo. Such problems clearly arose with regard to passage of TA96. For example, one of the issues that was perhaps the most difficult to resolve was the terms for lifting the interlata entry restriction from the Regional Bell Operating Companies (RBOCs). Competitive local exchange providers greatly feared the potential losses, whereas the RBOCs clearly expected tremendous benefits, from RBOC interlata entry.

Based on experience from both the electric and telecommunications industries, deregulatory efforts have manifested the controversies one would expect from pursuit of retrenchment policies. Furthermore, these retrenchment problems – that impose their own political feasibility constraints – were often in direct conflict with the economic viability constraints of providing electric and telecommunications services. Particularly instructive are the experiences from the legislative and FERC activities associated with the recent crisis in the electric industry of California and the Western region of the U.S., and the passage and implementation of TA96 for the telecommunications industry. In this regard, both journalists and scholars have identified numerous issues representing conflicts among the economic viability and political feasibility constraints of adopting and implementing deregulatory policies. Many of these issues relate to problems of rate restructuring.

Perhaps the most significant issue relates to the level of retail residential rates. For example, under the recent California deregulatory legislation, wholesale rates were permitted to fluctuate but retail residential rates were frozen at existing levels. This price freeze was itself one of the compromises necessary for initial passage of the state legislation. As prices in the wholesale market dramatically changed,[17] the effects of the gap between wholesale and frozen retail rates severely disrupted the provision of electricity services. Rolling blackouts were instituted, and losses for privately-owned electric utilities (providing the distribution facilities) mounted to such high levels that they were being driven into bankruptcy.

Government intervention of some form was critical to assure continuing provision of electricity services on a widespread and reliable basis. The most straightforward means for addressing the economic viability problem for the utilities was to repeal the retail rate freeze and permit all – including residential – retail rates to rise to the level necessary to ensure that utilities could cover their costs (both past and future) and earn a reasonable return. As one might expect, this option was advocated by economists, but was vehemently opposed by California politicians and consumers. For months the California legislature, Californian U.S. Senators, and California Governor Davis – supported by consumer advocates – blamed the crisis on power generators for price gouging, seeking refunds of their profits. They also insisted that the solution was for FERC to cap wholesale rates in order to maintain "just and reasonable rates" for residential customers. In response, the federal government claimed that the electricity crisis was a state problem. As customers continued to suffer rolling blackouts and the utilities' financial conditions continued to deteriorate, FERC finally agreed to provide some partial relief through limited price caps on interstate wholesale rates, and California politicians finally acceded to a limited increase in retail residential rates. However, by this time, California was also forced to accept the responsibility for bailing out the utilities, committing to issue over $10 billion in state bonds. Although necessarily paraphrased here and therefore unable to reflect all the nuances and tactics, the behavior of state and federal politicians during this crisis are examples of blame avoidance strategies. The experience in California shows, particularly when blame avoidance strategies are employed, how intractable the problem can be to simultaneously satisfy both economic viability and political feasibility constraints. After all, throughout this whole ordeal it was clear that the politicians were aware of the economic realities, as Gov. Davis was often quoted as saying "If I wanted to raise rates, I could have solved this problem in 20 minutes."[18]

---

[17] There are many factors that contributed to the changes in the wholesale market, such as large increases in natural gas prices, droughts affecting the supply of hydropower, the regulatory characteristics of the spot market, the legal barrier against utility use of long term contracts with suppliers, and environmental regulations impeding the construction of new generation plants. *See* papers and presentations presented at "The California Electricity Market Meltdown: The End of Deregulation," Conference held at AEI-Brookings Joint Center for Regulatory Studies, Washington, D.C. (March, 2001).

The passage and implementation of TA96 has also been riddled with retrenchment problems and institutional factors impeding rate rebalancing policy reform in the U.S. Cherry (2000) discusses these at length, and also provides a comparative analysis of rate rebalancing policy reform in the EU. With regard to residential retail rates, the legislative resolution for balancing the economic viability and political feasibility constraints of local exchange competition is provided in the statutory provisions of section 254 on universal service. To address retrenchment problems associated with consumer interests, section 254 retains – and slightly modifies by adding the term "affordable" – the traditional public utility obligation of "just, reasonable, and affordable rates" (section 254(b)(1)). This perpetuates the legacy of the just price doctrine. Section 254 also creates new statutory restrictions on rate restructuring in order to appease current beneficiaries under the traditional telecommunications rate structure. Section 254(g) contains provisions that prohibit deaveraging of rates for interstate interexchange services;

---

[18] See "California's Davis To Support Raising Electricity Rates," Wall State Journal, April 6, 2001, p. A2.

and section 254(b)(3) requires reasonably comparable rates for services provided in urban areas to low income customers and to those in rural, insular and high cost areas. [19]

Both the electricity and telecommunications industries in the U.S. are regulated through a dual jurisdictional government regime. As such, this also gives rise to retrenchment politics among units of government in the context of deregulatory policies. As previously discussed, some of the economic viability problems under deregulatory policies arise from conflicting, or at least insufficient coordination among, federal and state policies. Wholesale rates that are free to fluctuate with the market combined with frozen retail rates are a recipe for economic inviability when wholesale rates exceed the retail levels. This has been particularly problematic for the electricity industry in California, in which both the federal and state governments declined responsibility for solving the crisis. For the telecommunications industry, insufficient coordination in determining local exchange costs and pricing among the federal and state jurisdictions may pose financial viability of the affected carriers (Cherry, 2000a). Which government entity must yield its policy preferences to another to achieve the necessary compatibility is highly contentious. The jurisdictional battle at the heart of *AT&T v. Iowa Utilities Board (1999)* – in which the FCC and state commissions fought over which had the power to determine unbundling policy – is illustrative.

*5.1.2 Retrenchment Problems of Explicit Funding Mechanisms*

Second, in order to better ensure the economic viability of carriers subject to these price restrictions in a competitive local exchange market, section 254 establishes a policy for the FCC and the states to implement explicit universal service funding. This policy is based on a framework of identifying specific services to be supported, identifying customers eligible to benefit from support, quantifying the amount of support required, determining what carriers are eligible to receive support, and designing a mechanism for collecting funds from carriers that will be distributed to eligible carriers. Implementation of each of these elements of the framework by the FCC has created controversies (Cherry, 1998 & 2000b; Cherry & Wildman, 1999b; Cherry & Nystrom, 2000).

One contentious and repeatedly delayed element of implementation has been funding to support high cost areas. This is not surprising given that rural carriers and non-rural carriers – as well as the states in which they reside – have such differing interests. Given the historical policies of cost separations and high cost support for rural carriers through access charges, rural carriers (and the states in which they reside) traditionally receiving high cost support have the most to lose – and non-rural carriers (and the states in which they reside) traditionally receiving no high cost support have the most to gain – from possible retrenchment of high cost support (Cherry, 2000b). Thus far, implementation of high cost support funding still remains incomplete, although rural carriers traditionally receiving some form of high cost support continue to receive at least that level of support post-TA96.

[19] Interestingly, the EU directives do not contain comparable restrictions (Cherry, 2000b). In the electricity industry in the U.S., there has also been resistance to restructuring residential retail rates on a real-time pricing basis.

Perhaps the most striking example of blame avoidance strategies consists of the obfuscatory tactics utilized to confuse the causal links of policy change and its effects with regard to the collection mechanism of explicit funding. Throughout the statutory text of TA96 itself, Congress frequently used evasive and ambiguous terminology (Cherry, 2000b, p. 373). But, most notably, Congress refused to even use the term "tax", when it was clearly exercising its taxing power and attempting to delegate it to the FCC (Cherry & Nystrom, 2000). In this regard, the obfuscatory strategies continued post-TA96 by the FCC's attempts to hide the resultant impact on rate levels of implementing the "tax". For example, FCC Commissioner Harold Furchtgott-Roth issued lengthy dissents in two different FCC proceedings, explaining his frustration with the FCC's blame avoidance tactics in implementing the e-rate for schools and libraries under section 254(h)(1)(B). The first was in the Truth-in-Billing Order,[20] and the second in the release of a Notice of Inquiry on the effect of long distance prices on low-volume users.[21] The following is an excerpt from his dissent in the Notice of Inquiry:

> The Commission has engaged in a public relations campaign to convince the Washington political establishment that massive increases in the e-rate tax could be offset by access charge reductions and that the American consumer need not ever know about either the access charge reduction or the increased e-rate tax. In this way, the Commission can claim that its new tax is not responsible for increased rates.
>
> *From its inception, the Commission has attempted to conceal the e-rate tax from consumers. It has done so through a series of actions, both formal and informal, to coerce long distance companies into hiding the tax.* First, it employed behind-

---

[20] First Report and Order and Further Notice of Proposed Rulemaking, Truth-in-Billing and Billing Format, FCC 99-82, CC Docket 98-170 (May 11, 1999).

[21] Notice of Inquiry, In the Matter of Low-Volume Long-Distance Users, CC Docket No. 99-249, released July 20, 1999.

> the-scenes threats and pressures. When that was unsuccessful, the Commission made its threats public by adopting unconstitutional "truth-in-billing" rules ostensibly designed to penalize "deceptive" billing practices that, in fact, limit how long distance carriers may identify e-rate tax line items on their bills. Now, the Commission is unholstering its biggest threat of all: the power to re-regulate the long distance industry [by this Notice of Inquiry]. (Dissenting Statement of Commissioner Harold Furchtgott-Roth, Section IV, emphasis added)

As with the e-rate, the FCC has refused to require that carriers place a surcharge on customers' bills as a means of collecting the taxes resulting from the funding of the high cost and low-income support programs as well – even though appropriate use of a surcharge would provide a more competitively neutral means of collecting the funds (Cherry, 1998). In fact, if carriers nonetheless seek to use a surcharge, the Truth-in-Billing Order restricts how carriers may describe the purpose of the surcharge in order to shield the FCC (and Congress) from accountability for the tax. In his dissenting statement in the Truth-in-Billing proceeding, Commission Furchtgott-Roth claimed that such a restriction violated the First Amendment freedom of speech rights of the carriers.

The preceding retrenchment problems have been associated with initial adoption and implementation of new deregulatory policies. Fulfillment of the underlying objectives require that these polices remain in effect over time; however, the sustainability of these policies over time is uncertain. As previous experience with welfare-related programs have shown, the more universalistic the programs, the more politically stable they are over time. In fact, continuing resistance to raise retail rates for the entire class of residential customers was a critical component contributing to the development of the electricity crisis in California. For this reason, one would expect the more residualistic programs to be less stable over time. Of the telecommunications universal service programs, the support for low income customers is therefore probably the most vulnerable to retrenchment in the future – not only is the number of beneficiaries smaller under this program, but, due to more modest financial resources of the beneficiaries, well organized forces in opposition are less likely to develop.

### 5.2. Expansion of the Public Utility Welfare State

While retrenchment efforts of deregulatory policies require renegotiation of the traditional public utility welfare state through increased reliance on rate rebalancing and explicit funding sources, there are also pressures to expand the welfare state in the telecommunications industry. These pressures arise in two distinct, yet interrelated contexts.

First, rapidly changing communications technologies are creating pressures to expand the scope of telecommunications services that are deemed essential to all Americans. This type of expansion can be viewed as an extension of what constitutes a business affected with a public interest under the common law.[22] In particular, several provisions of TA96 call for the expansion of telecommunications infrastructure to provide advanced telecommunications and information services. For example, one of the principles of universal service mandated by Congress is that "access to advanced telecommunications and information services should be provided in all regions of the nation" (section 254(b)(2)); and universal service is to be an evolving level of telecommunications services as technology and societal needs change (section 254(c)(1)(A)-(D). Furthermore, in section 706 of TA96, Congress declared that the FCC and state commissions are to encourage the deployment of advanced telecommunications capability to all Americans.

Second, changes in communications technologies also provide new means to deliver services traditionally recognized as part of the welfare state. In this regard, section 254(h)(1) requires delivery of telecommunications services at special rates to certain educational institutions and health care providers. In addition, in section 706, Congress expressly states that elementary and secondary schools and classrooms are to have access advanced telecommunications infrastructure. Thus, when implementing section 254(h)(1)(B), the FCC included access to the Internet as one of the elements of universal service to be provided to educational institutions. This inclusion was unsuccessfully challenged by some of the telecommunications carriers in the courts,[23] and accounts for a substantial portion of the annual $2 billion costs of funding the universal service support program for educational institutions and libraries. Another example is the requirement in section 254(h)(1)(A) that health care providers serving rural areas be charged rates for telecommunications services that are reasonably comparably to those charged to persons residing in urban areas. The FCC has also implemented this provision to provide universal service support for telecommunications services provided to such health care providers.[24]

---

[22] Recall from the discussion in Section 2.2, quoting the U.S. Supreme Court in *Wolff Packing Co. v. Court of Industrial Relations of Kansas* (1923), that the third category of businesses affected with a public interest are those which may not have been such at their inception but become so over time.

[23] *Texas Office of Public Utility Counsel v. FCC,* 183 F. 3d 393 (5th Cir. 1999), *cert. denied,* 120 S. Ct. 2212 (2000). Cherry and Nystrom (2000, pp. 125-129) discuss why the court's decision was flawed on this point.

[24] Carriers receive a credit against contributions otherwise owed to fund universal service programs when providing the required lower rates to health service providers serving rural areas.

The pressures of these telecommunications policies to extend social welfare are further complicating the ability to address the retrenchment problems described in Section 5.1 by posing further economic viability and political feasibility constraints. Expanding the scope of what telecommunications services are viewed as essential is significantly increasing the funding burden of universal service support mechanisms. This increases the financial pressures on carriers to pass through the tax – levied on them to fund the mechanisms – through surcharges on customers' bills.[25] Yet, such surcharges intensify politicians' efforts to utilize blame avoidance strategies in order to avoid accountability for the price increases needed to fund these programs.

## 6. Summary and Conclusions

The adoption and implementation of deregulatory policies for public utilities that are likely to fulfill the underlying objectives require the satisfaction of both political feasibility and economic viability constraints. However, closing the gap between these two types of constraints to achieve sustainable public utility policies is a challenging endeavor. For economic viability, policy choices must support private investment in the market as a general matter, and be compatible with the economic viability needs of the firms or industry subject to specific regulation. For political feasibility, policy choices must support the legitimacy of government itself, enable initial adoption of the policy, and be sustainable over time. Furthermore, these requirements cannot be evaluated in isolation but are interrelated, adding to the complexity of the task. In some cases, political feasibility constraints may require sacrifice of some economic efficiency objectives, or economic viability constraints may require modification or even abandonment of some political objectives. For these reasons, it is important not only for policymakers to better understand the economic realities that limit achievability of policy goals, but also for all parties attempting to influence the policy process to be aware of the political constraints that limit policymakers' choices.

For deregulatory policies in public utility industries in the U.S., the legacy of common law doctrines must not be overlooked. The medieval concept of the "just price" in economic exchange – evolving through the development of common law doctrines of "public callings" and "businesses affected with a public interest" – became a critical component of public utility regulation. The public utility obligation to charge "just and reasonable rates" has been codified in federal and state statutes and remains in force today. The prior policy choices embedded in the common law doctrines contribute to policy path dependence, creating inertia and daunting obstacles to restructuring of retail rates in the electric and telecommunications industries.

---

[25] Carriers argue that the tax is too large for shareholders alone to bear the cost, and that, without some form of price increase, they would also have more difficulty attracting investors.

In terms of understanding the political ramifications of attempts at retrenchment, the characteristics that traditional public utility regulation shares with other forms of welfare state regulation must also not be ignored. Traditional public utility obligations are heavily value-laden, with their origins in philosophical concepts of both commutative and distributive justice. Due to the dual, contractual and status-based nature of its obligations, public utilities function as privatized social welfare bureaucracies in providing public utility services to the general public. To the extent that all residential customers benefit from special rates, the regulation is universalistic in nature. To the extent that more narrowly targeted groups of customers – such as low income customers and those residing in high cost areas – benefit from special pricing programs, the regulation is residualistic in nature.

Experience with the difficulties in retrenching from other forms of welfare state regulation better enables us to understand the barriers faced by advocates of public utility deregulatory policies. As a general matter, universalistic policies are politically difficult to change; residualistic policies, once adopted, are more vulnerable to subsequent efforts of retrenchment. As expected, the economic pressures to retrench from government imposition of low residential retail rates has met with furious political resistance. Such resistance has contributed to the development and continuance of the electricity crisis in California, and required federal enactment in TA96 of statutory limitations on the restructuring of telecommunications rates. Additional political difficulties for the telecommunications industry arise from the retrenchment problems associated with establishing explicit universal service funding mechanisms and the blame avoidance strategies that policymakers' feel they must employ to hide the tax imposed on carriers. These, in turn, are exacerbated by the political pressures to expand the public utility welfare state through deployment of telecommunications infrastructure to provide advanced services to all Americans and to enhance the provision of traditional welfare state services through universal service funding of telecommunications serviced provided to educational institutions and health care providers serving rural areas. By their very nature, the more residualistic universal service programs established under section 254 are also likely to be vulnerable over time to future retrenchment efforts.

Failure to come to terms with the interrelationship of political feasibility and economic viability problems means that policies may be adopted that are worse than the status quo. Furthermore, the adverse consequences may be of potentially catastrophic proportions, as evidenced by the electricity crisis in California. Thus, the cost-benefit analysis of deregulatory policies requires a new calculus. The societal costs may not justify the attempt to harness all the economic efficiency gains (whether static or dynamic) expected from deregulatory policies, particularly in light of the greater political and economic stability that has historically been achievable through more traditional public utility regulation. Furthermore, the importance of stability in the provision of public utility services is also likely to be of increasing value in a refreshed cost-benefit analysis given the magnitude of the costs being incurred during the recent downturn in the technology sector of the U.S. and worldwide economies.

## References

AT&T v. Iowa Utilities Board, 525 U.S. 366 (1999).

Adler, E. 1914. "Business jurisprudence," 28 *Harvard Law Review* 135.

Baldwin, J. 1959. "The medieval theories of the just price," in *Transactions of the American Philosophical Society*, Vol. 49, Part 4. Philadelphia, PA: The American Philosophical Society.

Barnes, I. 1942. *The Economics of Public Utility Regulation*. New York, NY: F.S. Crofts & Co.

Black, A. 1993. "The juristic origins of social contract theory," *History of Political Thought*, 14(1), 57-76.

Bonbright, J. 1961. *Principles of Public Utility Rates*. New York, NY: Columbia University Press.

"The California Electricity Market Meltdown: The End of Deregulation," Conference held at AEI-Brookings Joint Center for Regulatory Studies, Washington, D.C. (March, 2001).

"California's Davis to support raising electricity rates," *Wall Street Journal,* (April 6, 2001), p. A2.

Cherry, B. 2001. "Filling the political feasibility and economic viability gap to achieve sustainable telecommunications policy," presented at the Sixth Asia Pacific Regional Conference of the International Telecommunications Society, Hong Kong (July, 2001).

Cherry, B. 2000a. "An institutional perspective on assessing real option values in telecommunications cost models. In J. Alleman & E. Noam (eds.), *Real Options: The New Investment Theory and Its Implications for Telecommunications Economics*. Norwell, MA: Kluwer Academic Publishers.

Cherry, B. 2000b. "The irony of telecommunications deregulation: assessing the role reversal in U.S. and EU policy," in I. Vogelsang & B. Compaine (eds.), *The Internet Upheaval: Raising Questions and Seeking Answers in Communications Policy*, 355-385. The MIT Press.

Cherry, B. 1999. *The Crisis in Telecommunications Carrier Liability: Historical Regulatory Flaws and Recommended Reform*. Norwell, MA: Kluwer Academic Publishers.

Cherry, B. 1998. "Designing regulation to achieve universal service goals: unilateral or bilateral rules," in E. Bohlin & S. L. Levin (eds.), *Telecommunications Transformation: Technology, Strategy and Policy*. Amsterdam, The Netherlands: IOS Press.

Cherry, B., & Nystrom, D. (2000). "Universal service contributions: an unconstitutional delegation of taxing power," *L. Rev. M.S.U.-D.C.L.,* 2000(1), 107-138.

Cherry, B., & Wildman, S. (2000). "Preventing flawed communication policies by addressing constitutional principles," *L. Rev. M.S.U.-D.C.L.,* 2000(1), 55-105.

Cherry, B., & Wildman, S. (1999a). "Institutional endowment as foundation for regulatory performance and regime transitions: the role of the U.S. constitution in telecommunications regulation in the United States," *Telecommunications Policy*, 23, 607-623.

Cherry, B., & Wildman, S. (1999b). "Unilateral and bilateral rules: a framework for increasing competition while meeting universal service goals in telecommunications," in B. Cherry, S. Wildman, & A. Hammond (eds.), *Making Universal Service Policy: Enhancing the Process Through Multidisciplinary Evaluation*, 39-58. Mahwah, NJ: Lawrence Erlbaum Associates.

Cranston, R. 1985. *Legal Foundations of the Welfare State*. London, UK: Weidenfeld and Nicholson.

First Report and Order and Further Notice of Proposed Rulemaking, Truth-in-Billing and Billing Format, FCC 99-82, CC Docket 98-170 (May 11, 1999).

Glaeser, M. 1957. *Public Utilities in American Capitalism*. New York, NY: The Macmillan Co.

Goldberg. V. 1980. "Relational exchange," *American Behavioral Scientist,* 23, 337-352.

Grieve, W. & Levin, S. 1996. "Common Carriers, Public Utilities and Competition," *Industrial and Corporate Change*, 5(4), 993-1011.

Iowa Utilities Board. v. FCC, 525 U.S. 366 (1999).

Habermas, J. (1999). *Between Facts and Norms: Contributions to a Discourse Theory of Law and Democracy (third printing)*. The MIT Press.

Hall, P. (1986). *Governing the Economy: The Politics of State Intervention in Britain and France*. Oxford University Press.

Hoepfl, H., & Thompson, M. (1979). "The history of contract as a motif in political thought," *The American Historical Review,* 84(4), 919-944.

Horwitz, M. 1977. *The Transformation of the American Law: 1780-1860.* Cambridge, MA: Harvard University Press.

Kingdon, J. 1995. *Agendas, Alternatives, and Public Policies (second edition).* HarperCollins College Publishers.

Langholm, O. 1998. *The Legal of Scholasticism in Economic Thought: Antecedents of Choice and Power.* New York, NY: Cambridge University Press.

Langholm, O. 1992. *Economics in the Medieval Schools.* The Netherlands: E.J. Brill.

Levy, B., & Spiller, P. (eds.) (1996). *Regulations, Institutions, and Commitment: Comparative Studies of Telecommunications.* Cambridge University Press.

Mishra, R. (1990). *The Welfare State in Capitalist Society: Policies of Retrenchment and Maintenance in Europe, North America and Australia.* University of Toronto Press.

Munn v. Illinois, 94 U.S. 113 (1876).

Nebbia v. New York, 291 U.S. 502 (1934).

Notice of Inquiry, In the Matter of Low-Volume Long-Distance Users, CC Docket No. 99-249. released July 20, 1999.

Panzar, J. & Wildman, S. 1995. "Network competition and the provision of universal service," Industrial and Corporate Change," *Industrial and Corporate Change,* 4, 711-719.

Pierson, P. (1994). *Dismantling the Welfare State: Reagan, Thatcher, and the Politics of Retrenchment.* Cambridge University Press.

Pinch, S. 1997. *Worlds of Welfare.* New York, NY: Routledge.

Schumpeter, J. 1994. *History of Economic Analysis* (reprinted with a new introduction by M. Perlman). New York, NY: Oxford University Press.

Skocpol, T. (2000). *The Missing Middle*. W. W. Norton & Company, Inc.

Skocpol, T. (1995). *Social Policy in the United States*. Princeton University Press.

Texas Office of Public Utility Counsel v. FCC, 183 F. 3d 393 (5th Cir. 1999), cert. denied, 120 S. Ct. 2212 (2000).

Twight, C. 1991. "From claiming credit to avoiding blame: the evolution of congressional strategy for asbestos management," *Journal of Public Policy,* 11(2), 153-186.

Weaver, R. K. (1986). "The politics of blame avoidance," *Journal of Public Policy* 6(4), 371-398.

Wilsford, D. (1994). "Path dependency on why history makes it difficult but not impossible to reform health care systems in a big way," *Journal of Public Policy,* 14(3), 251-283.

Wilson, W. (1987). *The Truly Disadvantaged*. University of Chicago Press.

Wolfe, A. 1989. *Whose Keeper? Social Science and Moral Obligation*. University of California Press.

Wolff Packing Co. v. Court of Industrial Relations of Kansas, 262 U.S. 522 (1923).